\documentclass[pra,twocolumn,showpacs,superscriptaddress,floatfix,nofootinbib]{revtex4-2}

\usepackage{color}
\usepackage{hyperref}  
\usepackage{graphicx}
\usepackage{amsfonts}
\usepackage{footmisc}

\usepackage{fancyhdr}
\let\a=\alpha \let\b=\beta     
  \let\h=\eta     \let\l=\lambda
\let\m=\mu    \let\n=\nu             \let\r=\rho

 \let\D=\Delta   
     \let\F=\Phi

\font\tenmib=cmmib10\font\sevenmib=cmmib7\font\fivemib=cmmib5%
\mathchardef\Bl   = "0515  

\def\Bl   {{\mbox{\boldmath$ \lambda$}}}

\def\BDpr {{\mbox{\boldmath$ \partial$}}}

\def\eqalign#1{\null\,\vcenter{\openup\jot
  \ialign{\strut\hfil$\displaystyle{##}$&$\displaystyle{{}##}$\hfil
      \crcr#1\crcr}}\,}

\def\AA{{\mathcal A}}
\def\EE{{\mathcal E}}\def\DD{{\mathcal D}}
\def\MM{{\mathcal M}}

\def\uu{{\V u}}\def\kk{{\V k}}\def\xx{{\V x}}

\def\T#1{{#1_{\kern-3pt\lower7pt\hbox{$\widetilde{}$}}\kern3pt}}

\def\ie{{\it i.e.\ }}
\def\dpr{{\partial}}
\def\defi{{\buildrel def\over=}}

\def\otto{\,{\kern-1.truept\leftarrow\kern-5.truept\to\kern-1.truept}\,}
\def\Pprod{\prod^{\kern-1mm\raise.0mm\hbox{$\leftarrow$}}}
  
\newdimen\xshift \newdimen\xwidth \newdimen\yshift \newdimen\ywidth

\def\ins#1#2#3{\vbox to0pt{\kern-#2pt\hbox{\kern#1pt #3}\vss}\nointerlineskip}

\def\eqfig#1#2#3#4#5{
\par\xwidth=#1pt \xshift=\hsize \advance\xshift
by-\xwidth \divide\xshift by 2
\yshift=#2pt \divide\yshift by 2
{\hglue\xshift \vbox to #2pt{\vfil
#3 \includegraphics{#4.eps}
}\hfill\raise\yshift\hbox{#5}}}

\def\V#1{{\bf #1}}
\def\lis#1{{\overline#1}}

\def\tende#1{\,\vtop{\ialign{##\crcr\rightarrowfill\crcr
 \noalign{\kern-1pt\nointerlineskip} \hskip3.pt${\scriptstyle
   #1}$\hskip3.pt\crcr}}\,}

\def\0{\noindent}
\def\*{\vskip2mm}
\def\media#1{\langle #1 \rangle}
\def\Eq#1{\label{#1}}
\def\equ#1{(\ref{#1})}

\font\titolo=cmbx12%
\def\iniz{\setcounter{equation}{0}}
\def\be{\begin{equation}}\def\ee{\end{equation}}

\def\alertb#1{{\color{blue}#1}}

\newcounter{appendice}

\def\alert#1{{\color{ired}#1}}
\definecolor{iblue}{RGB}{65,105,225}
\definecolor{ired}{RGB}{220,20,60}
\definecolor{igreen}{RGB}{50,205,50}
\definecolor{ipurple}{RGB}{75,0,130}
\definecolor{iochre}{RGB}{218,165,32}
\definecolor{iteal}{RGB}{51,204,204} 
\definecolor{imauve}{RGB}{204,51,153}

\def\ap{{\it a priori }}
\iniz

\begin{document}

\alert{\centerline{\titolo Reversibility, Irreversibility,
    Friction and}}
\alert{\centerline{\titolo nonequilibrium ensembles in N-S equations}}
\vskip1mm
\centerline{\bf Giovanni Gallavotti\footnote{\small Posta-e:
  giovanni.gallavotti@roma1.infn.it,\\  Web: https://ipparco.roma1.infn.it}
}
\centerline{Universit\`a ``La Sapienza'' e INFN, Roma}
\centerline{\today}

{\vskip3mm}
\noindent {\bf Abstract}: {\it Viscosity, as a physical property
of fluids, reflects an average effect over a chaotic microscopic
motion described by Hamiltonian equations. It is proposed, as an
example, that stationary states of an incompressible fluid
subject to a constant force, can be described via several
ensembles, in strict analogy with equilibrium Statistcal
Mechanics.}{\vskip3mm}

\def\SEC{Regular NS evolutions}
\section{\SEC}
\label{sec1}
\def\Dot#1{{\bf\dot#1}}
\0{\bf Question:} can the phenomenological notion of friction be
represented in alternative ways?

\0{\bf Related Question:}  is it possible to set up a theory of
statistical ensembles, and their equivalence, extending to
stationary non-equilibria the ideas behind the canonical
and microcanonical ensembles, \cite{Ma867-b,SJ993}.

\*
A guide could be provided by the existence of a fundamental
symmetry like “time reversal” which cannot be “spontaneouly
broken”.\footnote{\small In the subatomic world time reversal is
not a symmetry but another more fundamental symmetry, CPT, could
replace it in the following discussion}

Therefore even the stationary states of dissipative systems ought
to be describable via time reversible equations.  Clearly the
question is not an easy one: hence it will be better to
specialize here on a paradigmatic example, namely the
Navier-Stokes (NS) fluid in a $2\pi$-periodic box, in $2$ or $3$
dimensions (2D or 3D), viscosity $\n$.

The $d$D equation, $d=2,3$, is:
\be\eqalign{
  \dot\uu(\xx)_a=&-((\uu\cdot\BDpr)\uu)_a-
  \BDpr_a p\cr
  &+\n\D\uu_a+{\bf F}_a=0, \quad \BDpr\cdot\uu=0\cr}\Eq{e1.1}\ee
where $\uu(\xx)$ is the velocity field that in 2D can be represented
via Fourier's series as:
\be \uu(\xx)_a = \sum_{\kk\ne\bf0} i u_{\kk} \frac{\kk^\perp_a}{||\kk||}
e^{-i\kk\cdot\xx}\Eq{e1.2}\ee
and $\bf F$ is likewise represented via its Fourier's series.

In
terms of the complex scalars $u_\kk=\lis\uu_{-\kk}$ the 2D NS
equation is:
\be \eqalign{\dot u_\kk =& -\sum_{\kk_1+\kk_2=\kk}
\frac{(\kk_1^\perp\cdot\kk_2)(\kk_2\cdot\kk)}{||\kk_1 ||\,||\kk_2 ||\,||\kk ||}
u_{\kk_1} u_{\kk_2}\cr&-\n\kk^2u_\kk+F_\kk\cr}
\Eq{e1.3}\ee

Although the 2D-NS admit general smooth solution $t\to S_u\uu$
starting from smooth initial data $\uu$, it is convenient (aiming
to discuss also the 3D-NS) to imagine them as truncated at
$|\kk|=\max_i|k_i|\le N$. The ultraviolet (UV) cut-off $N$ will be
temporarily fixed.  The 2D-NS become $M_N=(2N+1)^2-1$ dimensional
ODE’s, on phase space $\MM_N$.

In the 3D case the 3D-NS
equations with UV cut-off $N$ can be likewise written on a
$M_N=2((2N+1)^3-1)$ dimensional phase space $\MM_N$.

The time reversal transformation $I\uu=-\uu$ 
does not imply $IS_t\uu = S_{-t} I\uu$ if $\n>0$: hence these
are irreversible equations.

Let $\uu$ be an initial state: then $t\to S_t\uu\defi \uu(t)$ evolves and
it is easly seen that $||S_t\uu||$ will verify an \ap estimate on
$\|\uu(t)\|^2_2=\sum_\kk |u_\kk(t)|^2 $:
\be \|\uu(t)\|^2_2=\le
\|\uu(0)\|^2_2+\frac{\|\bf F\|^2_2}{\n^2}< \Big(\frac{\|\bf
  F\|^2_2}{\n^2}\Big)_+\Eq{e1.4} \ee
where the last inequality abridges stating that if
$\|\uu(0)\|^2_2$ is larger than $\frac{\|\bf F\|^2_2}{\n^2}$ then
it will decrease to become eventually, smaller than any target
$>\frac{\|\bf F\|^2_2}{\n^2}$ (in a time depending on
$\uu(0)$ and on the prefixed target).
\*

\0{\bf Definition:} {\sl A smooth dynamical system $S_t$ on a manifold $\MM$
will be called {\it regular} if in $\MM$ there are a finte number
of open sets $\AA_i, i=1,2,\ldots,n$ which:
\\
(1) are $S_t$-invariant, \\
(2) their union is $\MM$ up to a set of zero volume,
(3) the evolution of almost all data $\uu\in\AA_i$ assigns an
average value to the observables $O$ (\ie to functions on $\MM$)
$\m_i(O)$ where $\m_i$ is an invariant distribution.
\\
The $\m_i$ will be called ``physical distributions'' and if $n>1$
the system will be said to admit $n$ phases.}
\*

The distributions are called ``physical dustribution'' because
they allow to compute the time averages of functions $O(\uu)$ on
phase space: hence determine the statistical properties of almost
all data.

If the motion is chaotic then generates a ``stationary state'' on
$\MM_N$, \ie a stationary probability distribution which, aside
exceptions on the initial data collected in a $0$-volume set in
$\MM_N$, will be supposed unique, for simplicity, and denoted
$\m_\n (d\uu)$. 

Stationary probability distributions generalize to the case of
much more general systems, the regular ones,
what in many classes of Hamiltonian systems are the equilibrium distributions
studied in  equilibrium statistical mechanics
(SM).

Here we assume that the regularized NS equations, denoted
$INS^N$, on $\MM_N$ parameterized by the viscosity $\n$ are
regular in the above sense. And it is natural to define the
collection $\EE^{N}_{viscosity}$ of all physical distributions
$\m^N_\n (d\uu)$ on $\MM_N$. In the cases in which there are
several physical distributions corresponding to the same
viscosity a further label $\b=1,\ldots,n_\n^N$ will be attached as
$\m^N_{\n,j}$ to distinguish them.

Furthermore we fix once and for all the forcing $\F$, with $||\bf
F||_2=1$, and with $|F_\kk|=0$, if $|\kk|>k_{max}$ with
$k_max<\infty$: physically this is read that $\bf F$ is assumed
to be a large scale forcing.

\def\SEC{Enstrophy ensemble}
\section{\SEC}
\label{sec2}
\iniz

Consider the new equation, \cite{Ga996b}, with UV-cut-off $N$:

\be \eqalign{\dot u_\kk =& -\sum_{\kk_1+\kk_2=\kk}
\frac{(\kk_1^\perp\cdot\kk_2)(\kk_2\cdot\kk)}{||\kk_1 ||\,||\kk_2 ||\,||\kk ||}
u_{\kk_1} u_{\kk_2}\cr&-\a(\uu)\kk^2u_\kk+F_\kk, \qquad
|\kk|,|\kk_1|,|\kk_2|\le N\cr}
\Eq{e2.1}\ee
differing from the Eq.\equ{e1.3} because the multiplier $\a(\uu)$
replaces the viscosity $\n$. The non linear term in Eq.\equ{e2.1}
will be denoted ${\bf n}(\uu,\uu)_\kk$.

The multiplier $\a(\uu)$ will be so defined that the solutions of
the Eq.\equ{e2.1} will admidt
$\DD(\uu)\defi\sum_\kk\kk^2|u_\kk|^2$, usually called {\it enstrophy},
as an {\it exact constant of motion}. From Eq.\equ{e2.1} this means that

\be\a(\uu)=\frac{\sum_\kk \kk^2 \Big({\bf n}(\uu,\uu)_\kk+F_\kk
  \Big) \lis u_\kk}{\sum_\kk\kk^4|u_\kk|^2}\Eq{e2.2}\ee
(in 2D the term with $\bf n(\uu,\uu)$ vanishes identically).

The new equation is reversible: $IS_t \uu = S_{-t} I\uu$ (as
$\a(\uu)$ is odd) and will be called $RNS^N$.  So $\a$ can be
called a “reversible friction”.\footnote{\small In the 3D case
the equations are very similar (for instance the $u_\kk$ are
replaced bu vectors orthogonal to $\kk$, ..): the main and
important difference is that the expression of $\a(\uu)$ receives
a contribution from the quadratic transport term which, although
present also in 2D, cancels from $\a$ essentially because in 2D
when $\n=0$ the $\DD(\uu)$ is conserved.}

The non Newtonian forces $\n\D\uu$, $\a(\uu)\D\uu$ can be imagined to play
the role of a thermostat:\cite{Ga021} the forcing performs work on the fluid
but the fluid density remains the same. This means that the
temperature must change so that the equation of state linking
pressure density and temperature remains fulfilled. Heat has to
be removed or inserted and both forces
can be viewed as external forces allowing to achieve respect of
the equation of state. Then it is natural to think that the two
equations should be equivalent.

This leads to an equivalence conjecture, \cite{Ga020b}, which we
formulate still making use of our knowledge of Statistical
Mechanics where examples of equivalences are well known and can
guide us.

The evolution with $RNS^N$ wil be assumed regular in the sense of
the above definition; it generates a family of stationary by the
distributions on phase space: $\m^N_D$ parameterized by the
constant value $D$ of the enstrophy $\DD(\uu)$. Again for a given
$D$ there may be $m^N_D>1$ physical states which will be
distinguished by extra label $j=1,\ldots,n_D^N$. The collection
of such distributions will form the ``enstrophy ensemble'',
$\EE^N_{enstrophy}$.

To formulate the connection between the two ensembles
$\EE^N_{viscosity},\EE^N_{enstrophy}$ it is necessary do define
the {\it local observables} $O(\uu)$: these are functions on
phase space whose value depends on the harmonics $\kk$ contained in a
finite region $\D$ {\it independent} on $N$,\footnote{\small
Hence depend on finitely many Fourier's
harmonics of $\uu$, \ie are ``large scale'' observables'', but
unlike the forcing, which only has the harmonics $|\kk|<k_{max}$
with $k_{max}$ fixed, no limit is set on the size scale}.

The local observables here are analogous to the SM observables on
phase space depending only on the positions and velocities of
particles located in finite region $\D$ {\it independent} on the
size $V$ of the container. So once more arises a similarity
between the  SM of a system enclosed
in a volume $V$ and of a NS fluid  with UV cut-off $N$: one can
say that locality in SM is in position space while in NS it is in
momentum space.

With this in mind the following conjecture has been proposed to
relate physical distributions $\m^N_{\n,\b}, \r^N_{D,\b'}$ in
$\EE^N_{viscosity},\EE^N_{enstrophy}$ where $\b=1,\ldots,n^N_\n;
\b'=1,\ldots,m^N_D$ are labels distinguishing the physical
distributions, if more then one: note however that cases in which
$n^N_\n,m^N_D>1$ are expected to be rare. Then:\cite{Ga020b}
\*

{\bf Conjecture:} {\sl Let $\m^N_{\n,\b}\in\EE^N_{viscosity}$ and
$\r^N_{D,\b'}\in\EE^N_{enstrophy}$ be physical
distributions with the same enstrophy
\be\m^N_{\n,\b}(\DD)=D \Eq{e2.3}\ee
Then if $N$ is large enough it is
$n^N_\n=m^N_D$\ \footnote{\small
In many cases no ``intermittency'' is expected, \ie $n^N_\n=m^N_D=1$.}
and for each
$\b$ there is $\b'$ 
\be \lim_{N\to\infty} \n^N_{\n,\b}(O)
=\lim_{N\to\infty} \r^N_{D,\b'}(O)\Eq{e2.4}\ee
for all local observables.}
\*

So the averages of large scale observables will show the same
statistical properties, as $N\to\infty$ in $INS^N$ and $RNS^N$
evolutions, under the correspondence condition of equal
enstrophy, Eq.\equ{e2.3}.

The $\a(\uu)$ in the $RNS^N$ evolution will fluctuate strongly in
turbulence regime  and it will “self-average” to a constant $\n$
thus “homogenizing” the equation and turning it into the
$INS^N$ with friction $\n$.

The conjecture however does not mention a condition like for $\n$
small enough: it should hold also at high viscosity where often
$INS^N$ exhibits periodic attractors. The reason it is proposed
also in such cases is that in all cases the NS equations
should be regarded as yielding macroscopic descriptions of
microscopic, certainly chaotic, evolutions derived via scaling
limits without modifyig the microscopic equations, \cite{Le008}.

It is natural to think that there should be no condition for
strong chaos.  The microscopic motion is always strongly chaotic
and the chaoticity condition should be always fulfilled even when
motion appears laminar.

The analogy with SM becomes even more clear: the $N\to\infty$
limit corresponds to the thermodynamic limit $V\to\infty$.

As a final remark other ensembles can be imagined: first is the
``energy ensemble'' formed by the physical distributions for the
equation Eq.\equ{e2.1} with $\a$ so defined that the energu
$\|\uu\|^2_2$ is an exact constant (\ie $\a$ is given, in 2D and
also in 3D by Eq.\equ{e2.2} replacing $\kk^2$ with $1$ and
$\kk^4$ with $\kk^2$) and a similar analysis and conjecture be can be
set up: for results obtained with this approach see
\cite{SDNKT018}.

At this point it is convenient to pause and show a few
results from simulations which begin to test the equivalence
proposal.

\def\SEC{Some 2D simulations}
\section{\SEC}
\label{sec3}
\iniz

Here are collected results the earlier (<2018) simulations  that
might interest readers: skipping this part does not preclude
following the final comments in Sec.IV.

There is a first obvious test suggested by a rigorous consequence
of the conjecture based on the remark that the work per unit time
of the forcing is a local observable, being $W={\bf
  F}\cdot\uu\equiv \sum_{|\kk|<k_max} F_\kk\lis u_\kk$.
Multiplying both sides of $RNS^N$ or $INS^N$ by $\lis u_{\kk}$
one finds the energy conservation identity:

\be\eqalign{\a(\uu)D=& {\bf F}\cdot\uu,\qquad RNS^N\cr
  \n \DD(\uu)=& {\bf F}\cdot\uu,\qquad INS^N\cr}\Eq{e3.1}\ee

The conjecture implies that $\r^N_D(\a)D$ has to be equal in the
$\lim_{N\to\infty}$ to the average $W$ which in the $INS^N$ by the
  second line of Eq.\equ{e3.1} is $\n\m^N_\n(\DD)$. Hence in
  absence of intermittency the equivalence condition
  $\m^N_\n(\DD)=D$ yields
\be \lim_{N\to\infty} \r^N_D(\a)=\n\Eq{e3.2}\ee
Hence a test is: i) fix $\n,N$ and run the $INS^N$ evolution from a
random initial $\uu$ until the average enstrophy value $D$ of the
enstrophy $\DD(\uu(t))$ is numerically reached. The run the
$RNS^N$ with initial $\uu$ adjusted to have enstrophy $D$: the
result should be that if $N$ is large enough the running average
of $\a$,\ie $\frac1t\int_0^t\frac{\a(\uu(t'))}\n
dt'\tende{t\to\infty} 1$. An example is:

\eqfig{185}{140}{}{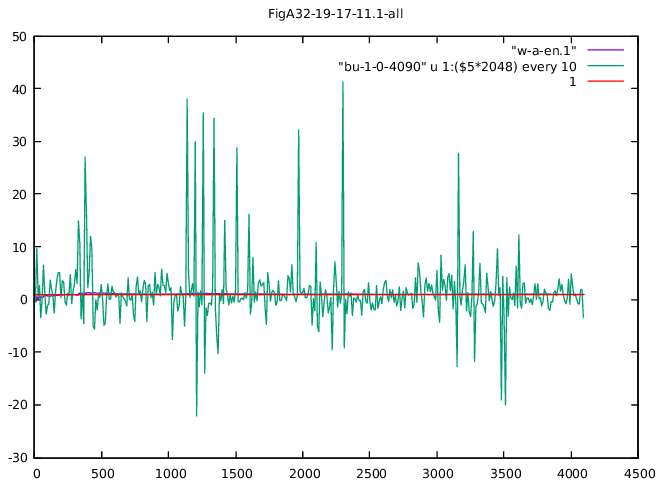}{
}

\0{\small Fig.1: The running average of the reversible friction
  $\frac{\a(\uu(t))}\n\equiv \frac1\n\frac{\sum_{\kk} \kk^2 2
    f_{-\kk} u_{\kk}} {\sum_\kk \kk^4|u_\kk|^2}$, superposed to
  the \alertb{conjectured value $1$} and to the fluctuating
  values $\frac{\a(\uu(t))}\n$: Evolution is by \alert{$RNS^N$},
  \alertb{\bf R=2048}, 224 modes ($N=7$), Lyap.-max $\simeq 2$,
  $x$-axis unit $2^{19}$, forcing only on modes
  $\kk=\pm(2,-1)$. Data are obtained via a sequence of
  integration steps of size $h=2^{-19}$ registered every
  $4h$. The plot gives $4000$ successive registered results but
  only every $10$ of them to avoid too dense a plot.}

\eqfig{185}{140}{}{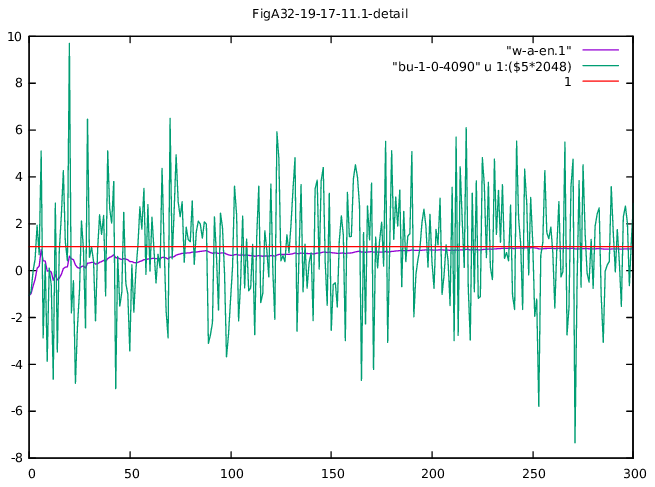}{}

\0{\small Fig.1-detail: The running average of the reversible friction
$\frac{\a(\uu(t))}\n$, superposed to the conjectured limit value
  $1$. Same data as the Fig.1 but for the shorter time interval
  $[0,300]$ to show the initial transient.}

The Fig.2 shows the preliminary evaluation of the average
enstrophy: it shows that in the case considered the average
avaleu of the enstrophy is reached quite rapidly in spite of the
strong fluctuations.

\eqfig{185}{140}{}{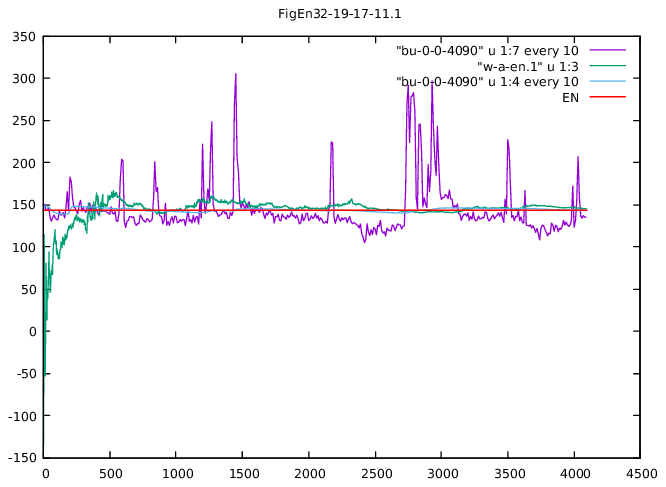}{}

\0{\small Fig.2: Running average of $\frac{W(\uu(t))}\n$ (dark green) in $INS^N$;\\
the average $\frac1\n D$ of $\frac1\n \DD(\uu)$ in $INS^N$ (straight red line)\\
running average of $\frac1\n \DD(\uu(t))$ in $INS^N$ 'converging'
to $\frac1\n D$ (very close to $\frac1\n D$)\\
large fluctuations are those of $\frac1\n \DD(\uu(t))$. Data are
the same in the previous figures.}

It is natural  to study the observable $\a(\uu)$ as it evolves
under the $INS^N$ equation. It is not covered by the conjecture,
that only implies that its average should be, under the  equivalence condition,
close to $1$ in the $RNS^N$ evolution.  An unexpected result is 
that the running average of $\frac1\n\a(\uu(t))$ also has running
average very close to $1$ as indicated by the following figure:

\eqfig{185}{140}{}{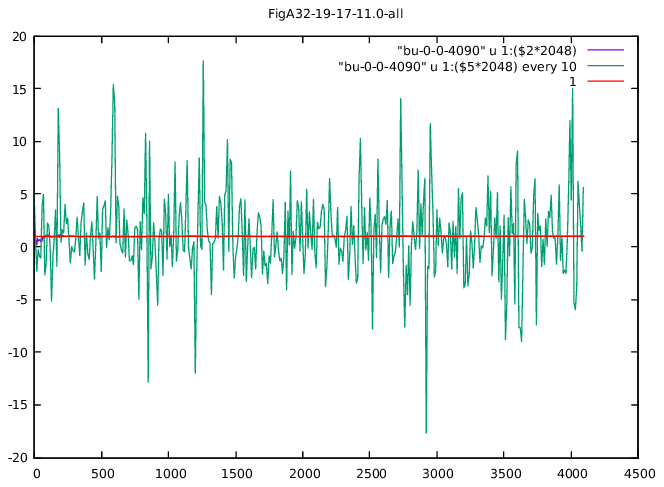}{}

\0{\small Fig.7: Same as Fig.1 but for $INS^N$.
The running average of the reversible friction
  $\frac{\a(\uu(t))}\n$ in a $N=7$ regularized $INS^N$ evolution
  forced on the mode $\kk=\pm(2,-1)$: the running average is in
  the large fluctuations curve. }

This is one more example of a non local observable with equal
averages in corresponding physical distributions for the two
evolutions considered.

The equality to $\n$ under the equivalence condition
between the average value of
$\a(\uu)=\frac{\sum_\kk \kk^2 F_\kk\lis u_\kk}{\sum_\kk
  \kk^4|u_\kk|^2}$  considered as an observable for both $RNS^N$
and $INS^N$ with $\n$ is perhaps surprising.
It is a theorem (consequence of the
conjecture) in the $RNS^N$ evolution but it might not be even expected
in the case of $INS^N$ because $\a(\uu)$ is not a local
observable.

Therefore it is tempting to test possible equality of the
averages of other nonlocal quantities, as such equalities are
well known to hold in the thermodynamic limit for several non
local observables.

One such observable is the spectrum of the symmetric part of the
Jacobian $J(\uu)$, the $M_N\times M_N$ matrix
$J_{\kk,\kk'}=\frac{\dpr\cdot u_\kk}{\dpr u_{\kk'}}$, 
  formally $\frac{\dpr \cdot\uu}{\dpr \uu}$. The spectrum can be
  called the ``local Lyapunov spectrum''.

\eqfig{185}{140}{}{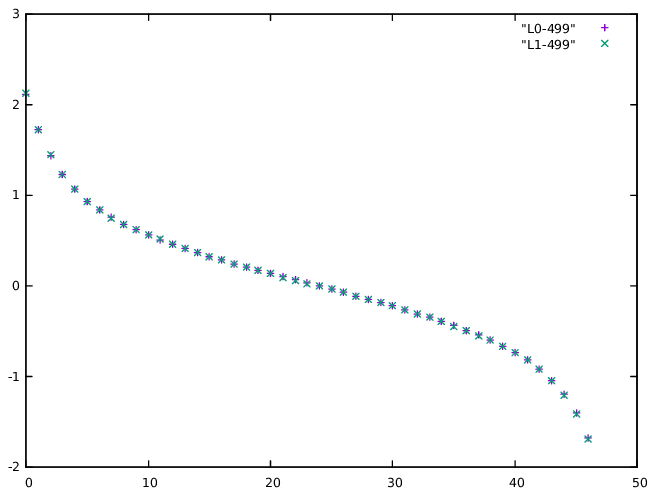}{}

\0{\small Fig.3: The spectrum of the symmetrized Jacobian called the
``local Lyapunov exponents'' for $N=3$ ($48$ modes) truncation for
$INS^N$ and $RNS^N$ under the equivalence condition: they are
superposed but undistinguishable on the scale of the picture
which plots $(k,\l_k)$, same data as in the previous pictures
(but truncated at $N=3$).}

The above coincidence to some extent is due to the wide ordinate
scale used. It is expected the two spectra are subject to
computatinal errors which should be more visible near the $k$ to
which correspond $\l_k$'s close to $0$. This is clarified in

\eqfig{185}{140}{}{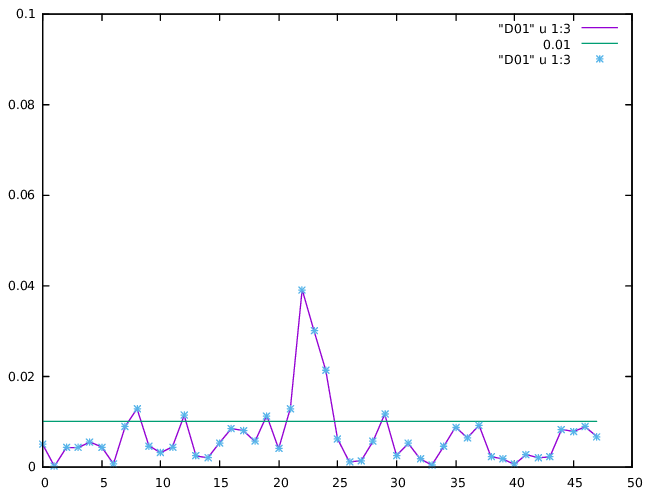}{}

\0{\small Fig.4: Relative difference betweeen (local) Lyapunov exponents
  in the previous Fig.3 ($\n=2048^{-1}$, $N=3$ (\ie $48$ modes)).}

Certainly a cut-off at $N=3$ is much too small to be of any
significance. In fact agreement between the $INS^N$ and $RNS^N$
is expected also at fixed $N$ and small $\n$, \cite{Ga997}, as a
consequence of appearance of turbulence: which generates
apparently random fluctuations on $\a$ with a corresponding
homogenization phenomenon: and physically diferent phenomenon.

The following figure tests the equivalence in the case of a
higher UV cut-off:

\eqfig{185}{140}{}{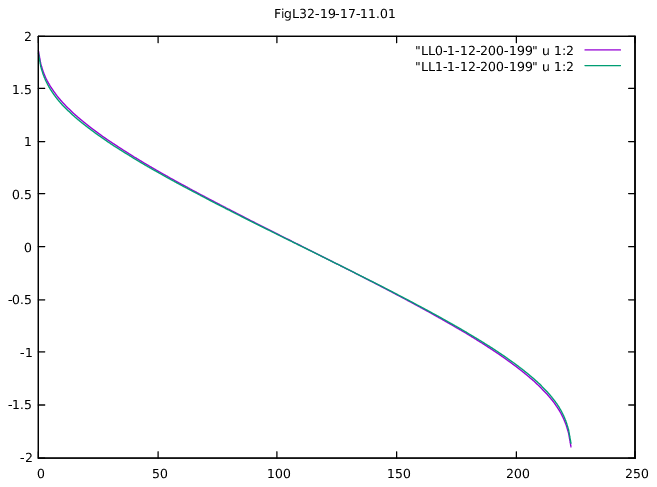}{}
  
\0{\small Fig.5: Local Lyapunov spectra in a $15 \times 15$
  ($960$ modes) truncation for $INS^N$ and $RNS^N$ (keys ending
  respectively in 0 or 1), with the $\l_k$ interpolated versus
  $k$ by lines,   $\n=\frac1{2048}$, forcing onthe
  modes$\kk=(2,-1)$. Each of the  $\l_k(\uu(t))$ is evalueted every
  $2^{19}$ integration steps and the graf reports the average of
  of each $\l_k(\uu(t))$ over $2200$ successive evaluations.}

The latter graph shows the spectra for both $INS^N$ and $RNS^N$
cases: again they are superposed. As in the previous case the
relative difference can be studied more clesely:

\eqfig{185}{140}{}{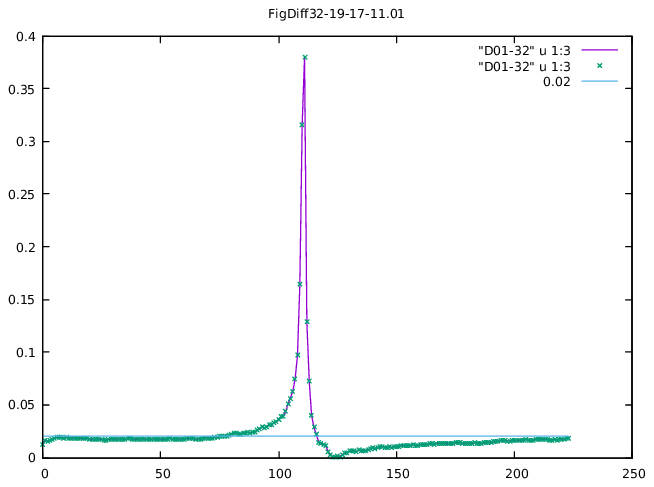}{}

\0{\small Fig.6: Relative difference betweeen (local) Lyapunov exponents
  in the previous Fig.5}

\def\SEC{Further results and problems}
\section{\SEC}
\label{sec4}
\iniz

{\bf(1)} The question of exhibiting examples of ``regular systems'' in the
sense of Sec.II has been essentially answered in the proposal by
Ruelle that I intepret as saying that ``generically'' all systems
exhibiting chaotic evolution should be ``regular''.

In more mathematically oriented works the idea emerges from the
theory of Anosov flows: they play the role, in chaotic dynamics,
of the harmonic oscillators in ordered dynamicsl they are the
paradigm of Chaos.  This idea rests on fundamental works of Sinai
(on Anosov sys.), and Ruelle, Bowen (on Axioms A systems).  A
strict, general, heuristic, interpretation of original ideas on
turbulence phenomena, \cite{Si968a,BR975,Bo975,Ru989,Ru978b}, led
to the, \cite{GC995b}:
\*
\0{\bf Chaotic hypothesis:} {\it A chaotic evolution takes place on a
smooth surface $A$, “attracting surface”, contained in phase
space, and on $A$ the maps $S$ (or the flow $S_t$) is an Anosov
map (or flow).}

So a regular system is a system with $n\ge1 $ attractors $A_i$ whose
basins of attraction are open sets $\AA_i$ whose union of the
entire phase space up to a set of zero volume.

\* {\bf(2)} The cahotic hypothesis is dismissed (by many) with
arguments like (1999) {\it More recently Gallavotti and Cohen
  have emphasized the “nice” properties of Anosov systems. Rather
  than finding realistic Anosov examples they have instead
  promoted their “Chaotic Hypothesis”: if a system behaved “like”
  a [wildly unphysical but well-understood] time reversible
  Anosov system there would be simple and appealing consequences,
  of exactly the kind mentioned above. Whether or not
  speculations concerning such hypothetical Anosov systems are an
  aid or a hindrance to understanding seems to be an aesthetic
  question.}\cite{HG012}

While giving up any evaluation of the statement I stress that
Statistical Mechanics, after Clausius, Boltzmann and
Maxwell was a simple and appealing consequence of the
“[wildly unphysical but well-understood]” periodicity
of motions of atoms in a gas.
\*

{\bf(3)} The same tests mentioned here (dated up to 2018) have been
made in 2D NS with up to $920$ modes and in 3D NS even up to
$\sim 5\cdot10^6$ modes.\cite{Ga021}

{\bf(4)} The 3D tests have suggested that the notion of local
observable should be made more strict defining $O$ a local observable
if $O(\uu)$ depends finitely many harmonics $\kk$, as above, further
verifying $|\kk|< c K_\n$
where $K_\n=(\frac\h{\n^3})^{\frac14})$ is Kolmogorov's scale and
$c$ is a constant and $\h=\n\media{W}$ is the average work per
unit time: it is believed widely the in the limit $\n\to0$ $\h$
should remain finite and positive.\cite{MBCGL022}

Although the 3D tests seem to suggest $c$ is of order $O(1)$ in
my view the possibility that $c=\infty$ should be studied with
further simulations.

{\bf(5)} one of the consequences of the chaotic hypothesis is that if
the system is time reversible then a general result is the
``fluctuation theorem'' (FT). The problem is that there is strong
evidence that in the NS the asymptotic motion is attracted on a
set of dimension lower than that of phase space: this is shown by
the fact that as soon as $N$ is large enough and $\n$ small
enough the attractor has dimension less that that of phase space:
this is shown by the $>0$ Lyapunov exponents seem to be less than
half the number of negative ones.\cite{Ga020b}

In 2D this seems to be the case with $\n=1/2048$ and $N=7$ and
$15$. There are a few examples in which even though the attractor
$A$ has dimension lower than that of phase space the motion on
$A$ admits a symmetry $I':A\to A$ which has the same properties
as time reversal (\ie $I'S_t=S_{-t}I'$): but is is unclear that
the NS admit such a symmetry. Positive results in testing
validity of the FT have been found, \cite{Ga020}, in 2D in the
$RNS^N$ with $N=48, \n=1/2048$, way too small for being really
interesting, while already for the case $N=224, \n=1/2048$ it is
unclear if FT holds.
\*

\def\Alert{}\def\Alertb{}
\centerline{\bf Appendix: A path through the theme}
\*
\01) \Alert{A first equivalence example:} \cite{SJ993}\\
2) \Alertb{Path to the conjecture:} \cite{Ga997b,Ga019c,Ga020b,MBCGL022}\\
3) \Alert{3D enstrophy ensemble:} \cite{MBCGL022,JC021}\\
4) \Alertb{3D energy ensemble:} \cite{SDNKT018}\\
5) \Alert{Shell model:} \cite{BCDGL018}\\
7) \Alertb{Stat-Mech:} \cite{Ru969,Ru977,Ru978b,Ru989,Ru012}\\
8) \Alert{Turbulence physics:} \cite{Fr995,BF010,Ge013,BV019,Fe000}

\*
\0{Acknowledgements: This is a redacted and updated version of a
  talk at the {\it DinAmici} meeting on 21/Dec/2018 at the
  Accademia dei Lincei, Roma.}

\bibliographystyle{unsrt}
\bibliography{268.bbl}

\end{document}